\documentclass[%
 aip,
 amsmath,amssymb,
reprint,%
]{revtex4-2}
\usepackage[utf8]{inputenc}
\usepackage{graphicx}
\usepackage{dcolumn}
\usepackage{subcaption}

\usepackage{bm}
\usepackage{siunitx}

\usepackage[T1]{fontenc}
\usepackage{xcolor}

\usepackage[colorlinks=true, urlcolor=blue,linkcolor=blue,citecolor=blue, pdfborder={0 0 0}]{hyperref}
\usepackage{caption}

\begin{document}


\title{Optimization of the characteristics of a relativistic electron beam based on laser wake-field acceleration using a non-symmetric gas target profile} 





\author{D. Mancelli}\thanks{Authors to whom correspondence should be addressed}\email{dmancelli@hmu.gr}

\affiliation{Institute of Plasma Physics \& Lasers, University Research \& Innovation Centre, Hellenic Mediterranean University, 74100 Rethymno, Crete, Greece}
\affiliation{Department of Electronic Engineering, School of Engineering, Hellenic Mediterranean University, 73133 Chania, Crete, Greece} 

\author{G. Andrianaki}
\affiliation{Institute of Plasma Physics \& Lasers, University Research \& Innovation Centre, Hellenic Mediterranean University, 74100 Rethymno, Crete, Greece}
\affiliation{Department of Electronic Engineering, School of Engineering, Hellenic Mediterranean University, 73133 Chania, Crete, Greece} 

\affiliation{School of Production Engineering \& Management, Technical University of Crete, 73100 Chania, Greece}

\author{I. Tazes}
\affiliation{Institute of Plasma Physics \& Lasers, University Research \& Innovation Centre, Hellenic Mediterranean University, 74100 Rethymno, Crete, Greece}
\affiliation{Department of Electronic Engineering, School of Engineering, Hellenic Mediterranean University, 73133 Chania, Crete, Greece}

\author{C. Vlachos}
\affiliation{Institute of Plasma Physics \& Lasers, University Research \& Innovation Centre, Hellenic Mediterranean University, 74100 Rethymno, Crete, Greece}
\affiliation{Department of Electronic Engineering, School of Engineering, Hellenic Mediterranean University, 73133 Chania, Crete, Greece} 
\affiliation{University of Bordeaux-CNRS-CEA, Centre Lasers Intenses et Applications, UMR 5107, 33405 Talence, France}

\author{I. Fitilis}
\affiliation{Institute of Plasma Physics \& Lasers, University Research \& Innovation Centre, Hellenic Mediterranean University, 74100 Rethymno, Crete, Greece}
\affiliation{Department of Electronic Engineering, School of Engineering, Hellenic Mediterranean University, 73133 Chania, Crete, Greece}

\author{I. Nikolos}
\affiliation{School of Production Engineering \& Management, Technical University of Crete, 73100 Chania, Greece}

\author{M. Bakarezos}
\affiliation{Institute of Plasma Physics \& Lasers, University Research \& Innovation Centre, Hellenic Mediterranean University, 74100 Rethymno, Crete, Greece}
\affiliation{Physical Acoustics and Optoacoustics Laboratory, Department of Music Technology \& Acoustics, Hellenic Mediterranean University, 74100 Rethymnon, Greece}

\author{E. P. Benis}
\affiliation{Institute of Plasma Physics \& Lasers, University Research \& Innovation Centre, Hellenic Mediterranean University, 74100 Rethymno, Crete, Greece}
\affiliation{Department of Physics, University of Ioannina, 45110 Ioannina, Greece}

\author{V. Dimitriou}
\affiliation{Institute of Plasma Physics \& Lasers, University Research \& Innovation Centre, Hellenic Mediterranean University, 74100 Rethymno, Crete, Greece}
\affiliation{Physical Acoustics and Optoacoustics Laboratory, Department of Music Technology \& Acoustics, Hellenic Mediterranean University, 74100 Rethymnon, Greece}

\author{N. A. Papadogiannis}
\affiliation{Institute of Plasma Physics \& Lasers, University Research \& Innovation Centre, Hellenic Mediterranean University, 74100 Rethymno, Crete, Greece}
\affiliation{Physical Acoustics and Optoacoustics Laboratory, Department of Music Technology \& Acoustics, Hellenic Mediterranean University, 74100 Rethymnon, Greece}

\author{M. Tatarakis}
\affiliation{Institute of Plasma Physics \& Lasers, University Research \& Innovation Centre, Hellenic Mediterranean University, 74100 Rethymno, Crete, Greece}
\affiliation{Department of Electronic Engineering, School of Engineering, Hellenic Mediterranean University, 73133 Chania, Crete, Greece}



\date{\today}

\begin{abstract}

We demonstrate a high-energy, high-charge, electron source produced by the irradiation of a novel gaseous target by an ultra-intense femtosecond laser pulse. By exploiting a nonsymmetrical nozzle, we increased the total charge of the electron beam by at least an order of magnitude with respect to our previous experiments using symmetrical nozzles. In addition, the maximum energy of the accelerated electrons was enhanced by a factor of two. The electrons are accelerated via the Laser Wake-Field Acceleration mechanism. Particle-in-cell simulations indicate that electrons are injected via the ionization and the downramp injection mechanisms. Our measurements indicate that the demonstrated electron source is a considerable candidate for high dose, Very High Energy Electrons applications, such as radiotherapy.
\end{abstract}


\maketitle 

\section{Introduction}\label{intro}
Particle acceleration is a research field that has been developing continuously for more than a century. The scientific and practical interest in charged particles and photon beams is wide; thus, huge investments have pushed the limits of accelerators’ capabilities and potential. However, conventional accelerators are progressing by creating larger facilities, where longer acceleration length is allowed, increasing their size and consequent cost in values that are under criticism \cite{accelerator_cost}.

Laser technology has developed rapidly and, together with plasma science, provides an alternative, as Laser Plasma Accelerators (LPA) have achieved acceleration of high-energy particle beams in millimeter to centimeter distances by creating accelerating fields exceeding several orders of magnitude larger than those created by radiofrequency cavities. LPA are extensively studied either as standalone accelerators \cite{record}, or in combination with a previously accelerated beam injected in a Laser Wake-Field for a second acceleration stage \cite{two_stage}. Their reduction in size and cost is their main advantage, while the most important drawback is their stability/reproducibility \cite{Mangles}. As the accelerator is practically recreated per shot, the resulting beams present fluctuating characteristics (e.g. energy, energy spread, total charge, and pointing), which limit their potential applications. In addition, Laser Wake-Field Acceleration (LWFA) \cite{Geddes2004, Faure2004,mangles_2004} is a method that simultaneously provides an electron and X-ray source, as oscillating electrons emit coherent betatron-type radiation in the keV range \cite{Rousse_betatron,Grigoriadis_APL,ong_2024}. TerraWatt (TW) and PetaWatt (PW) lasers, hosted in university-scale facilities, are currently working to improve beam characteristics and gain control and reproducibility \cite{Maier_100.000}.

LWFA accelerates relativistic beams with \SI{}{pC} \cite{pC} to \SI{}{nC} \cite{nC} total charge and energy up to a few \SI{}{GeV} \cite{record_Aniculaesei,record}. LWFA occurs when a high-intensity, ultra-short laser pulse propagates through an under-dense, gaseous target profile \cite{science.273.5274.472,Mo_LWFA2012}. The pulse drives plasma waves along its propagation through the homogeneous ionized gas. Typically, in a high-intensity laser, the intensity of its pre-pulse already exceeds the ionization threshold of electrons of a hydrogen or helium gas; thus, after the pre-pulse propagation, the gas has already evolved into a plasma. The ponderomotive force of the main part of the laser pulse pushes electrons away from the area where the electric field is maximum, namely along the propagation axis, creating an ion bubble that matches the dimensions of the local laser beam size, which changes rapidly due to self-focusing and self-steepening effects. The electrons oscillate under the influence of the space charge force and form plasma waves, with an electron density-dependent plasma wavelength
$\lambda_p [\mu m]=33 \times(n_e [10^{18} cm^{-3} ])^{-1⁄2}$ , which follow the laser pulse up to its gradual vanishment. 
Under specific conditions, the wave amplitude increases and breaks at the rear side of the bubble, allowing the injection of energetic electrons into the bubble. This electron injection mechanism, known as self-injection \cite{Lu_theory}, is mainly responsible for the creation of low-energy-spread electron bunches. However, other injection mechanisms may occur, such as ionization injection, where a multielectron gas is employed, either stand-alone or in a mixture with hydrogen or helium. In this case, the inner electrons of the multielectron gas hold an ionization threshold exceeded only by the peak of the main part of the laser pulse or even by the laser intensity reached when self-focused. These electrons are ionized into the bubble and gain energy, resulting in a more populated bunch with high energy and high-energy spread \cite{ionization,ionization_injection2,Tomassini2022,Grigoriadis_APL}. Another injection mechanism is the Downramp injection, which occurs when the laser pulse propagates along a density drop, commonly referred to as a downramp or density transition \cite{Kamperidis_Papp_2021,Massimo_2017,PIC_trancuated_ion_inj_Maity_2024,parabolic_optically_shaped,Hansson_2015, blade_cfd}. The ion bubble along the density drop is longitudinally elongated, permitting the injection of lower-energy electrons at its rear side. 
Finally, acceleration of the injected electrons takes place along the laser propagation direction, following an inverse plasma density law. Acceleration is limited by several reasons, such as laser pulse diffraction or depletion. The length over which these are happening, affected by self-guiding effects, is a few mm for a TW laser and extends up to few cm for PW lasers. Another limitation is dephasing, where electrons enter the deceleration phase of the field and lose energy, a tunable limit as it is related to plasma density.

In TW facilities, the basic setup for LWFA consists of the optics for focusing the laser pulse down to a few micron-radius focal spots on a gas jet emanating from a pulsed gas valve. More sophisticated setups involving gas cells/capillaries are usually employed in PW lasers, where the laser pulse is depleted and diffracted over longer distances. In addition, when the downramp mechanism is studied, the gas flow is tailored via several techniques. When the scale length of the density drop is sharp, namely comparable to the plasma wavelength, often referred to as shock front injection, the density profile is achieved primarily by the use of a blade that partially interrupts the gas jet, resulting in the development of a shock, which is a short and abrupt increase in the density of the flow density \cite{blade_cfd,Scchmid_2010,Kaluza_blade,PRL_blade}. Another setup related to downramp injection is by optically shaping the plasma density, pre-forming a lower density channel where injection is provoked \cite{parabolic_optically_shaped}. On the other hand, a longer density drop is created mainly by implementing consecutive nozzles that produce an intersecting gas area of higher density, followed by a lower density area \cite{Hansson_2015,Golovin_consequtive,Tomkus2020_multijet}. The aforementioned setups provide some tunability over electron beam characteristics via controlling blade or nozzle positions, or nozzle backing pressure separately, and enhance reproducibility \cite{Hansson_2015}. However, complex and expensive equipment (e.g., translation stages and/or extra gas valves) is required \cite{Tomkus2020_multijet}. In addition, due to the small size of the interaction area, the additional equipment is hard to establish in some configurations. 

In this context, a novel idea is the use of a single non-symmetrical nozzle to create a downramp profile, which, to the best of our knowledge, has been so far realized by only one group \cite{asymetrical}.
Our research team at the Institute of Plasma Physics and Lasers (IPPL) of the University Research \& Innovation Centre of the Hellenic Mediterranean University (HMU) has performed systematic studies on electron acceleration using the LWFA method \cite{Grigoriadis_2022,grigoriadis_scientific_reports,Grigoriadis_PPCF_2023,Grigoriadis_APL,Clarkal._2021}. In this work, we focus on the enhancement of the electron beam total charge with a goal of achieving a high-dose source. A high-dose source is preferable for use in biomedical applications and to generate efficient betatron radiation \cite{Ferri_betatron, labate_toward_2020}. The gaseous target is formed using, for the first time in our LWFA experiments, a novel in-house 3D-printed, non-symmetrical (NS) nozzle, designed to generate a unique gaseous density profile. The profile consists of a density downramp, followed by a quasi-constant density region aiming to enhance the downramp injection mechanism. The design of this nozzle removes the need for extra targetry features, thus reducing the setup complexity and making it favourable for applications. Following our previous work, we used the \SI{45}{TW} Ti:Sa Zeus laser system for the irradiation of N$_2$ density profiles, thus combining several injection mechanisms, with the aim of maximizing the electron dose.

\section{Results and Discussion} \label{resu}

 \begin{figure*}[]
  \centering
    \includegraphics[width=\textwidth]{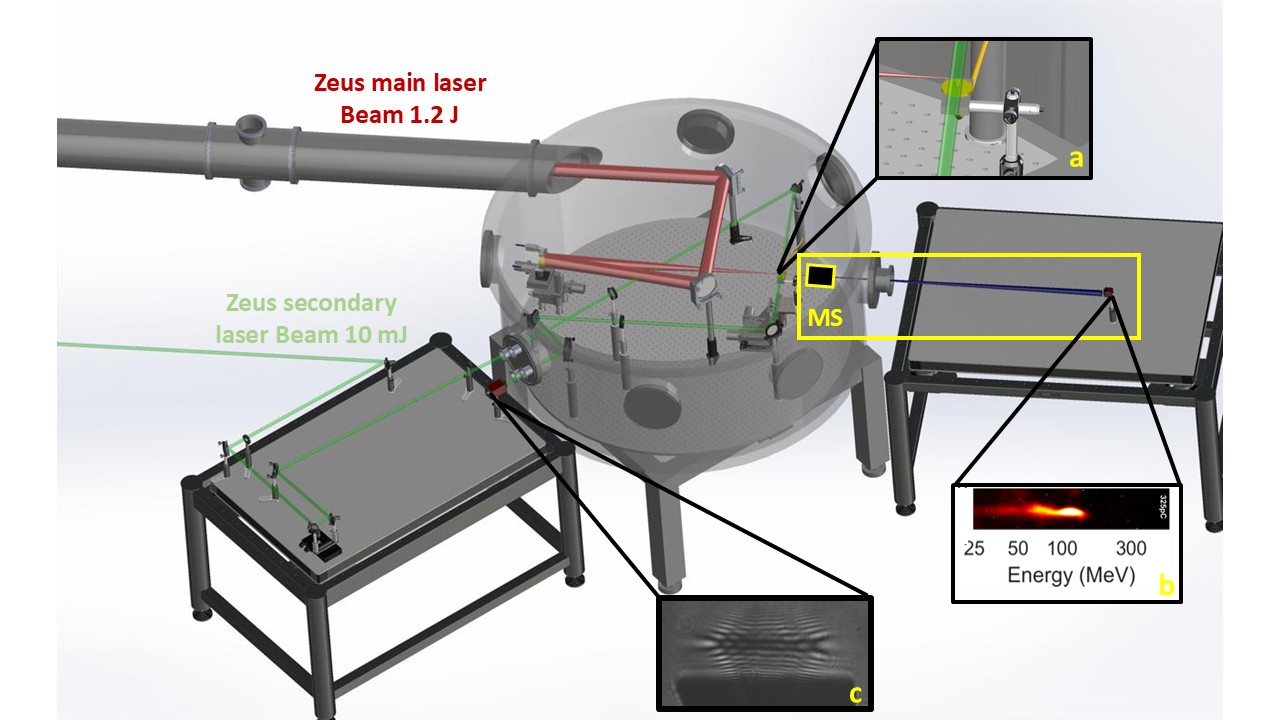}
\caption{The experimental setup. The main beam of the Zeus laser focuses downstream of the NS nozzle exit (a) and interacts with the gas. The accelerated electrons travelling in the laser propagation direction enter the magnetic spectrometer (MS), and their resulting spectra are recorded by a CCD (b). The low-energy, secondary beam serves to probe the interaction via shadowgraphy (c).}
\label{Fig:expSet}
\end{figure*}

The LWFA experimental setup layout is shown in Fig.~\ref{Fig:expSet} \cite{Grigoriadis_2022,clark_2021}. The main beam of the \SI{45}{TW}  Zeus laser system with a central wavelength at \SI{807}{\nano\meter} , is focused \SI{300}{\micro\meter} downstream of the exit of the nozzle. The focal spot of \SI{28}{ \micro\meter} diameter at Full Width Half Maximum (FWHM) is obtained using a \SI{1.008}{\m} focal length gold-coated off-axis parabola. The duration of the laser pulse is \SI{24}{fs}, resulting in a mean intensity of the order of \SI{e19}{W/cm^2} $(a_{0} \sim2.5)$. The NS nozzle is attached to an electromagnetic pulsed gas valve, triggered by the laser system to deliver the gas at tunable backing pressure. After the interaction, the accelerated electrons travelling in the laser propagation direction enter the magnetic spectrometer, which consists of two parallel magnetic plates with a magnetic field of \SI{0.3}{T} and a scintillating screen (Lanex Regular). Electrons are deflected according to their energy due to the magnetic field, and after crossing the screen, the light emitted is collected by a calibrated optical system (lens and CCD camera) \cite{Grigoriadis_phD}. The electron energy spectrum is calculated based on a home-developed algorithm \cite{Grigoriadis_phD}. The secondary \SI{10}{mJ} energy beam of Zeus laser system at the same laser wavelength and pulse duration is used as a probe beam for shadowgraphy measurements of the interaction. 
 
\begin{figure*}
     \centering
     \begin{subfigure}[b]{0.3\textwidth}
         \centering
         \includegraphics[width=\textwidth]{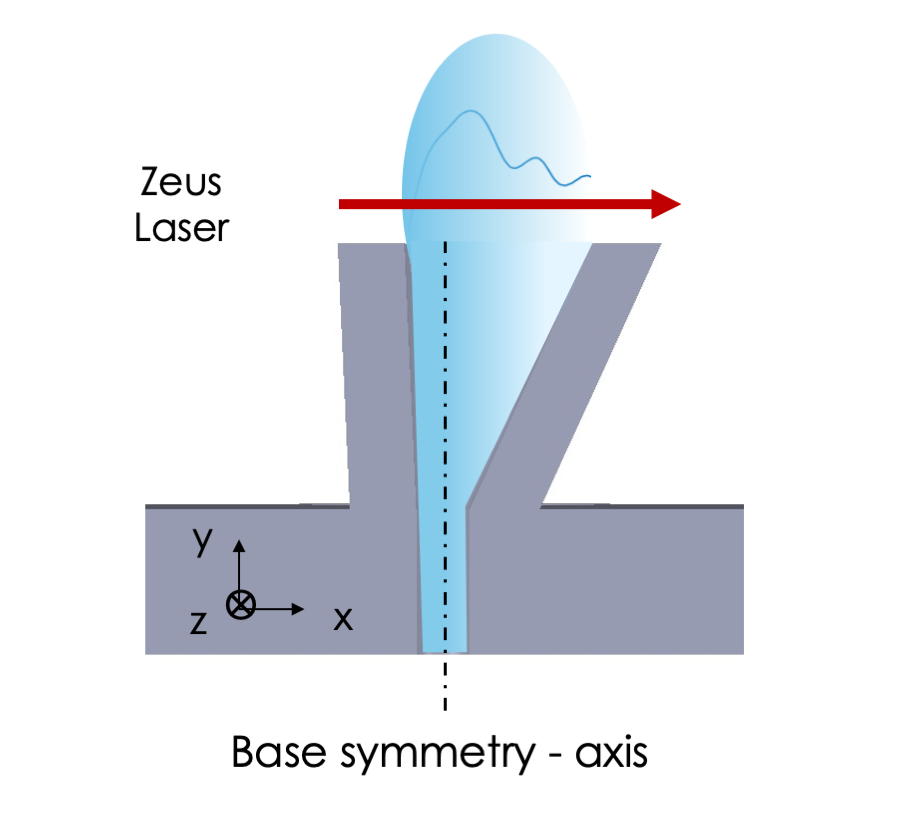}
          \caption{}
     \end{subfigure}
     \hfill
     \begin{subfigure}[b]{0.3\textwidth}
         \centering
         \includegraphics[width=\textwidth]{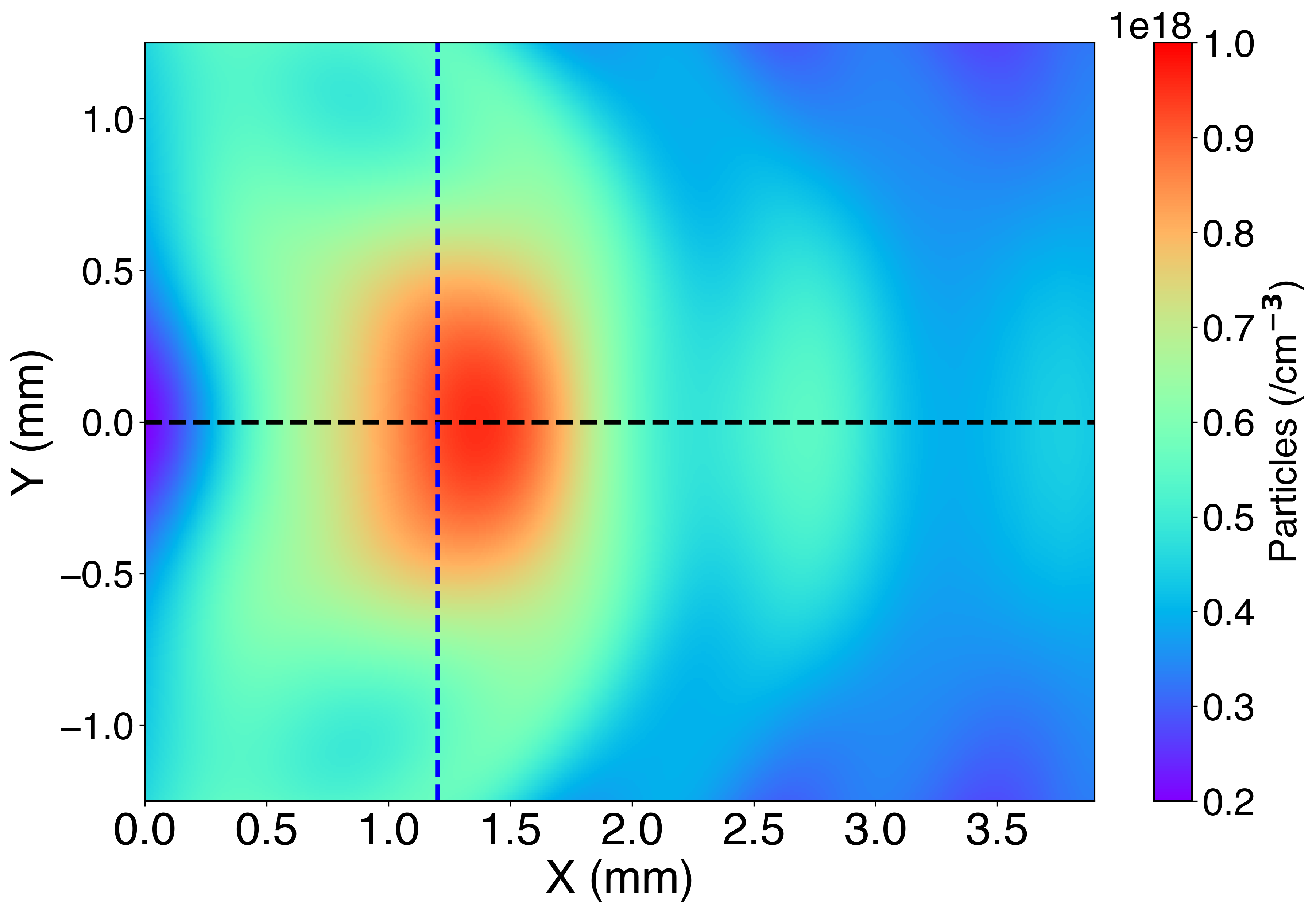}
         \caption{}
     \end{subfigure}
     \hfill
     \begin{subfigure}[b]{0.3\textwidth}
         \centering
         \includegraphics[width=\textwidth]{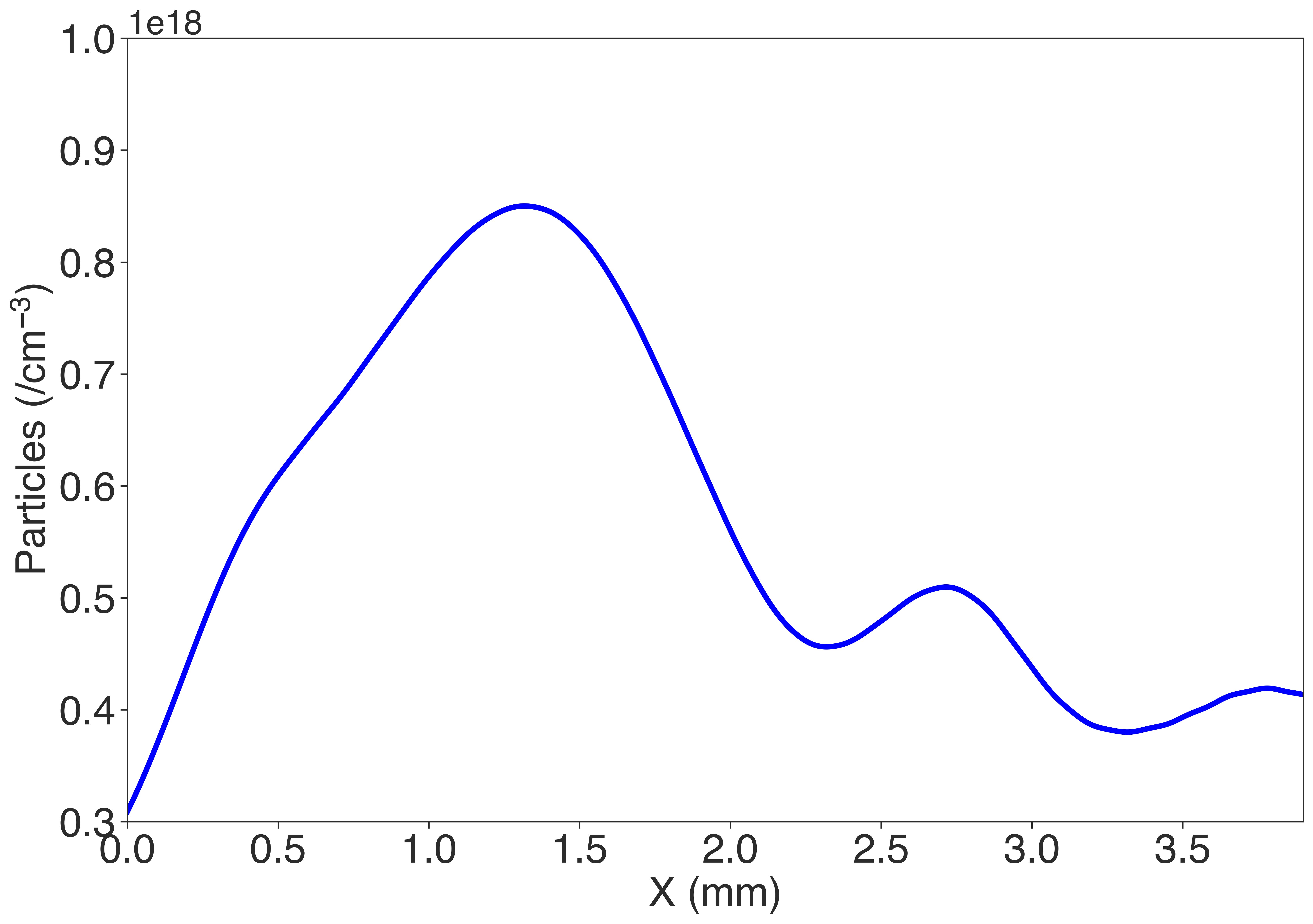}
         \caption{}
     \end{subfigure}
        \caption{(a) A CAD section view of the NS nozzle with a graph at the top indicating the produced gas density profile (blue line), along the laser propagation direction.  (b) Plane view of the particle density \SI{300}{\micro\meter} downstream the NS nozzle exit, obtained using a tomographic reconstruction method \cite{tomo_COUPERUS,HIPP200453}. (c) The density profile corresponding to the black dashed line of (b), namely the laser propagation axis at (Y=\SI{0}{mm}).}
        \label{Fig:jet}
\end{figure*}

The longitudinal length of the NS nozzle used in our experiments is \SI{3.5}{mm} and was manufactured via 3D printing \cite{Dopp_3d_jets,Vargas_3d_gas_cell,Andrianaki}. The nozzle was printed via stereolithography, using an ANYCUBIC printer with a resolution of \SI{0.05}{mm} on xy plane and \SI{0.01}{mm} along z-axis. Fig.~\ref{Fig:jet}(a) presents a cross-section of the nozzle CAD design, along with a qualitative density distribution of the gas flow. The throat is circular, with a diameter of \SI{800}{\micro\metre}, in acordance with the specifications of the gas valve at the connecting region. Toward the exit, it is mainly extended in one direction, forming an asymmetric ellipse, whereas the opposite side is flat. The Mach number corresponding to the elliptical part of the nozzle is M = 5.5 deduced from Computational Fluid Dynamic simulations. Fig.~\ref{Fig:jet}(b) presents a section view of the $N_2$ density \SI{300}{\micro\meter} downstream the nozzle exit obtained using tomographic reconstruction \cite{tomo_COUPERUS,HIPP200453} of multiple interferograms (for more details on the method applied see \cite{Andrianaki,proc-EPS}). Fig.~\ref{Fig:jet}(c) shows the density line-out obtained along the dashed black line shown in Fig.~\ref{Fig:jet}(b), which coincides with the laser propagation axis. A high-density area is formed, with a peak $\sim$\SI{e18}{particles/cm^3}, which drops along a downward ramp within approximately \SI{500}{\micro\meter} length. For longer distances, an almost constant, lower-density area is evident. 
    

\begin{figure*}[tbp]
  \centering
    \includegraphics[width=\textwidth]{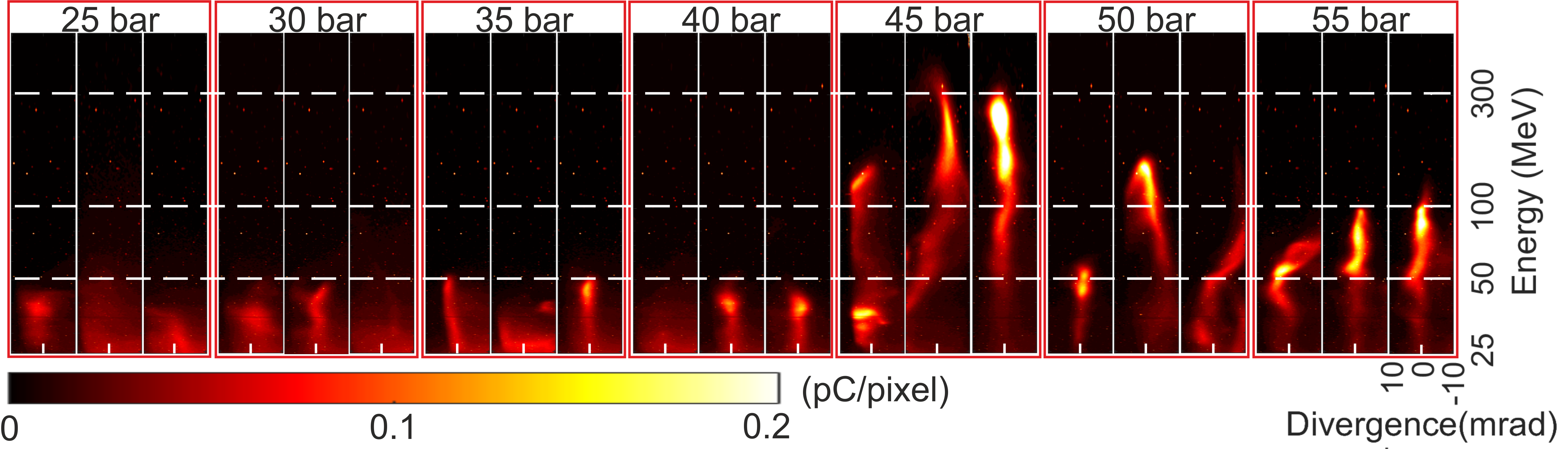}
\caption{A series of typical electron spectra measurements for different backing pressures at the gas valve inlet, ranging from \SI{25}{} to \SI{55}{bar}. Three representative shots per backing pressure are presented. The low energy limit of the spectrometer is \SI{25}{MeV}. The charge per pixel corresponds to the colour bar, and the angular divergences are displayed.}
\label{Fig.spec}
\end{figure*}

%
The interaction of the laser with the gaseous profile resulted in the generation of electron bunches with beam-like characteristics. The variation of the electron beam characteristics was examined by changing the backing pressure applied at the gas valve inlet. In Fig.~\ref{Fig.spec} characteristic shots are presented per backing pressure. The lower detection limit of the magnetic spectrometer was \SI{25}{MeV}. At the low-pressure regime, high-charge, low-energy electrons are seen to have a large divergence, measured with remarkable reproducibility at $\sim$\SI[separate-uncertainty = true]{20}{\milli\radian}.
By increasing backing pressure, the accelerated electrons improve their beam-like characteristics, having a lower divergence, while the electron energy increases and the total beam charge remains almost constant. The maximum energy is observed at a backing pressure of \SI{45}{bar} with the measured electron energy reaching \SI[separate-uncertainty = true]{300(20)}{MeV}, and divergence less than \SI[separate-uncertainty = true]{10}{\milli\radian}. An enhancement of the electron energy by at least a factor of two is achieved in comparison to our previous work.
In addition, we would like to stress that the charge/dE/shot was enhanced by at least a factor of $\sim$ 20 reaching a value of \SI{0.2}{(\pico\coulomb/pixel)} instead of \SI{0.01}{\pico\coulomb/pixel} \cite{Grigoriadis_PPCF_2023}, for N$_2$ gas, compared to our previous works where symmetrical nozzles were studied \cite{Grigoriadis_PPCF_2023,Grigoriadis_2022,Andrianaki}. The electron signal was observed in every shot, although variations in the energy spectra were evident as shown in Fig.~\ref{Fig.spec}. Thus, injection was ensured, and is attributed to the simultaneous occurrence of two injection mechanisms, downramp and ionization injection that take place during the interaction, as strongly suggested by Particle-In-Cell (PIC) simulations, presented below.

\begin{figure}[]
\centering
\includegraphics[width=\linewidth]{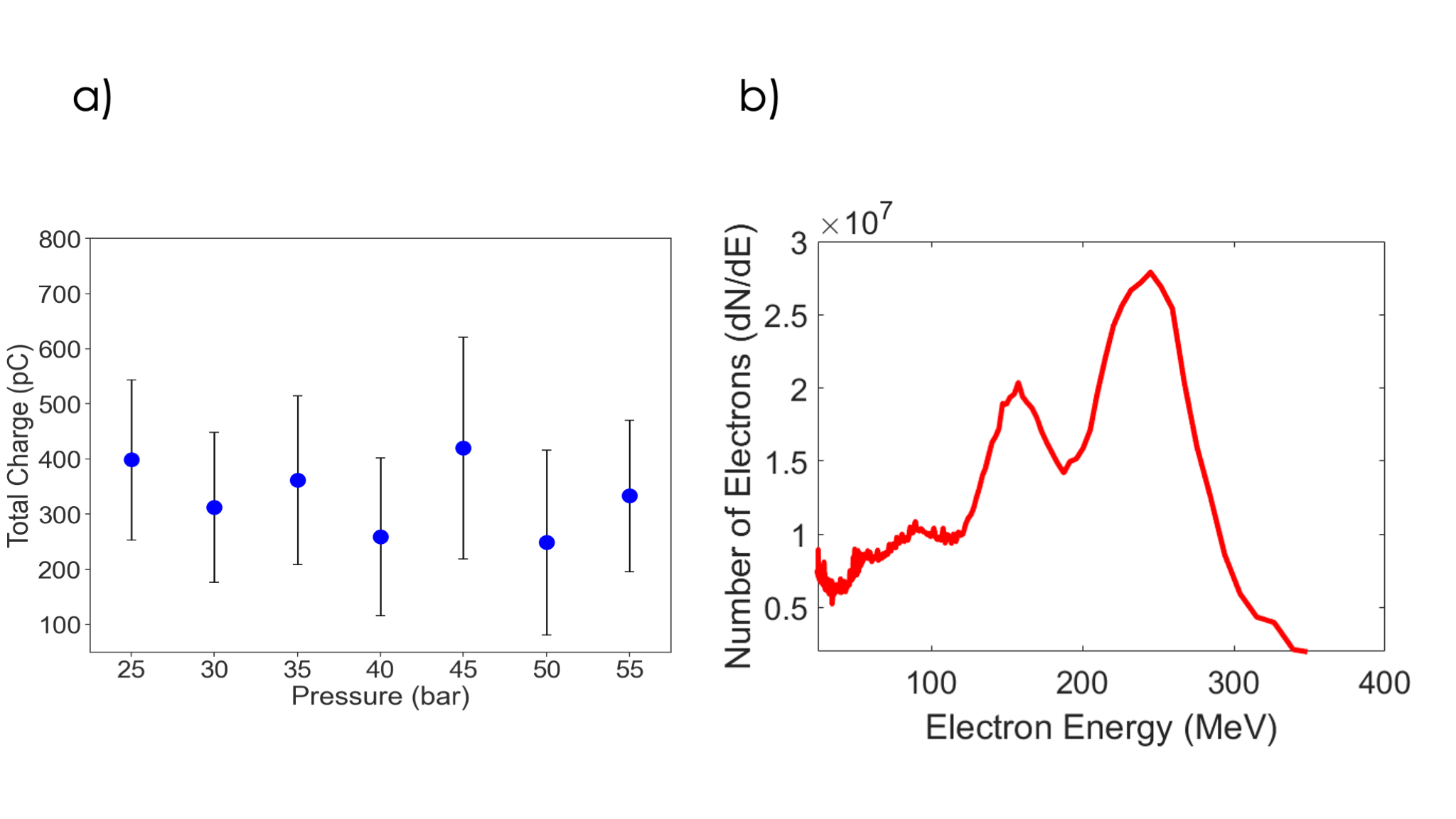}

\caption{(a) Total charge as a function of gas valve backing pressure. Measurements from \SI{17}{}-\SI{22}{} shots per backing pressure were averaged. (b) Energy spectrum of the highest energy shot at \SI{45}{bar} backing pressure.} 
\label{Fig.dose}

\end{figure}

\begin{figure}[]
    \centering
\includegraphics[width=\linewidth]{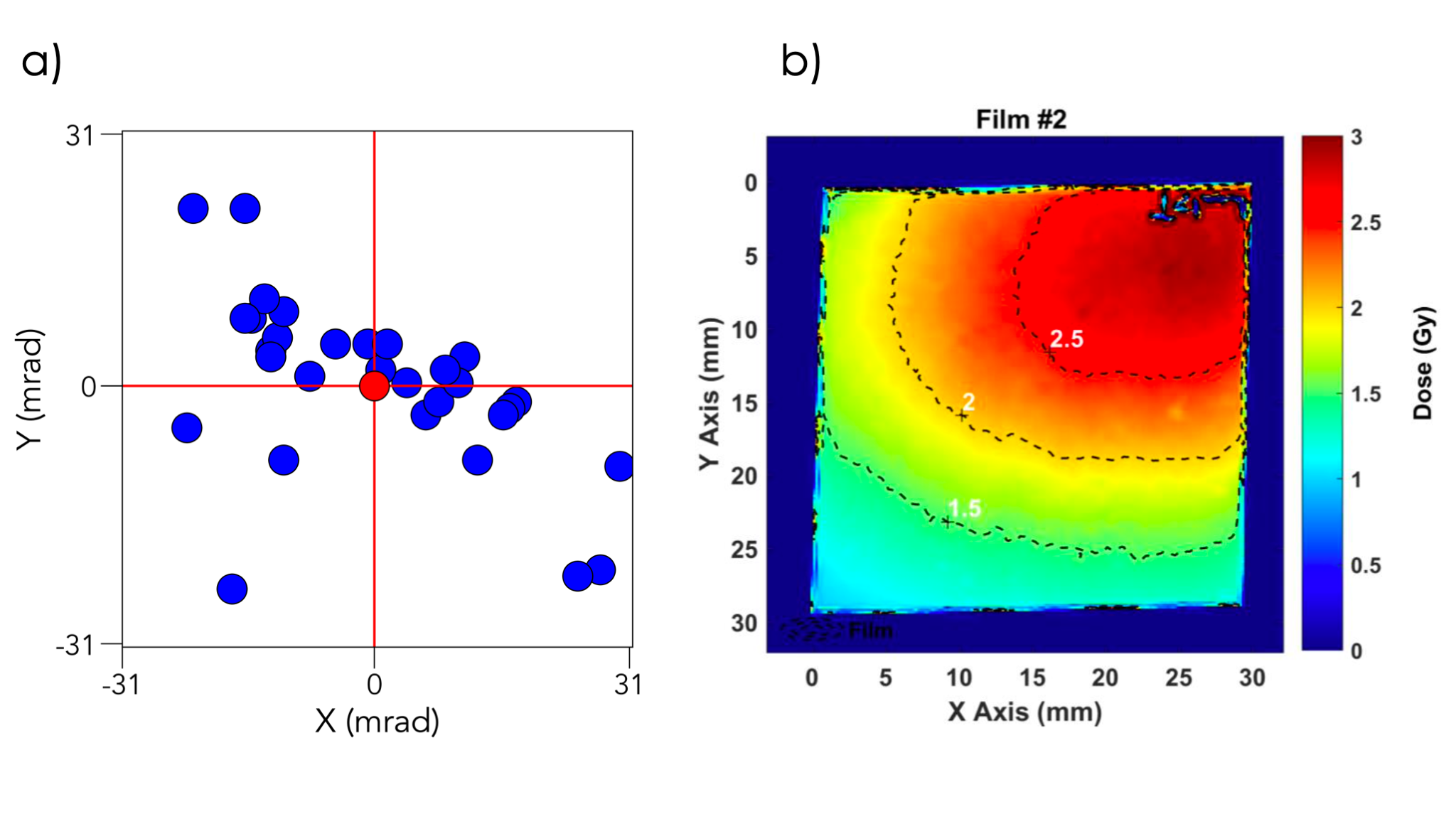}
\caption{(a) The beam pointing angle spread of the electron source (blue marks) of several consecutive shots. Red mark indicates the expected zero point and (b) 2D colour map of the measured absolute dose on the RCF Film aggregated over \SI{140}{} shots. A clear iso-dose distribution ranging from \SI{1.5}{Gy} the outer line to \SI{2.5}{Gy} the central iso-line localised on a region of \SI{15}{} $\times$ \SI{15}{mm} is shown.}
\label{fig:iso}
\end{figure}

Fig.~\ref{Fig.dose}(a) shows the median total charge, in red-filled circles, with the statistical error bars for several shots, as a function of gas valve backing pressure. The total charge of each shot is found by integration of the signal over the whole area displayed on the CCD camera. The total charge remains almost constant with respect to the backing pressure.

Fig.~\ref{Fig.dose}(b) shows the electron energy spectrum for the case of the highest energy and total charge obtained at the optimum operating backing pressure window, found to be between 45 and\SI{55}{bar}.

At all backing pressures there is a faint background (as, for instance is clearly shown in Fig.~\ref{Fig.spec} at the low-energy area of the spectrum) which in some shots is present along the whole detector area. As evident in Fig.~\ref{Fig.spec}, by increasing the backing pressure, this faint background is drastically reduced, and the accelerated electron bunches acquire sharper beamlike characteristics. Within the optimum
operating backing pressure window, the accelerator yields high energy, high-dose electron bunches with satisfying stability. The low-energy electron beams produced in the low-pressure regime are also useful, with the high electron population having energy \SI{25}{MeV}. This energy is similar to that of the electron beams used for radiotherapy with LINACs \cite{labate_toward_2020}.


The spread of the beam pointing angle of
the electron source for several consecutive
shots is shown in Fig.~\ref{fig:iso}(a). The electron dose distribution was measured and is shown in Fig.~\ref{fig:iso}(b). After removing the magnetic spectrometer, a super-sensitive, absolutely calibrated radiochromic film (EBT3 Gafchromic) was placed \SI{3}{cm} after the chamber electron exit port, normal to the electron beam propagation direction. The electron beam after passing through a \SI{2}{cm} glass polymerizes the detector's material. The minimum captured energy was \SI{6}{MeV} due to glass filtering \cite{nist-estar}.

The dose of 140 consecutive shots was accumulated. The accumulated dose results in a Gaussian-like distribution. It is worth noticing that at the central iso-surface region of about \SI{10}{mm} in diameter, a dose $\geq$\SI{2.5}{Gy} is measured. The mean dose per shot is $\geq$\SI{15}{mGy} which exceeds the values reported in previous results from experiments conducted under similar conditions \cite{Beyreuther_2010,Kokurewicz_2017}. When operating at \SI{5}{Hz} the source it may deliver to this central area a dose of $\ge$\SI{4.5}{Gy/min}. This value is adequate for experiments exploring the potential use of LWFA accelerated electrons in biomedical applications \cite{labate_toward_2020}.

\section{Two-dimensional PIC simulations}

\begin{figure}[]
\centering
\includegraphics[width=\linewidth]{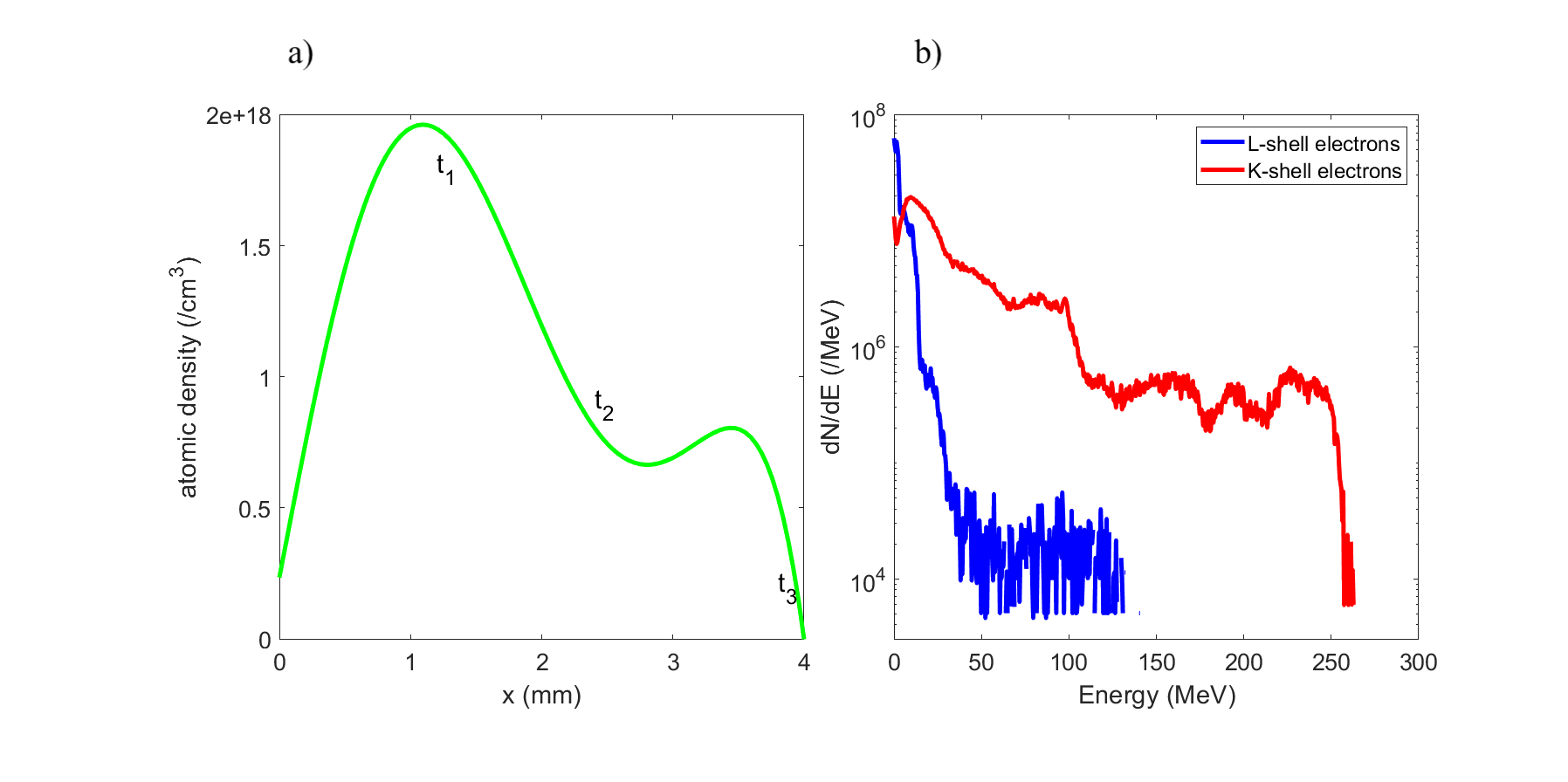}

\caption{(a) Polynomial approximation of atomic $N$ density used for the PIC simulations. (b) Electron energy spectrum resulting from PIC simulations}
\label{Fig.spec_sim}

\end{figure}

\begin{figure*}[]
  \centering
\includegraphics[width=\textwidth]{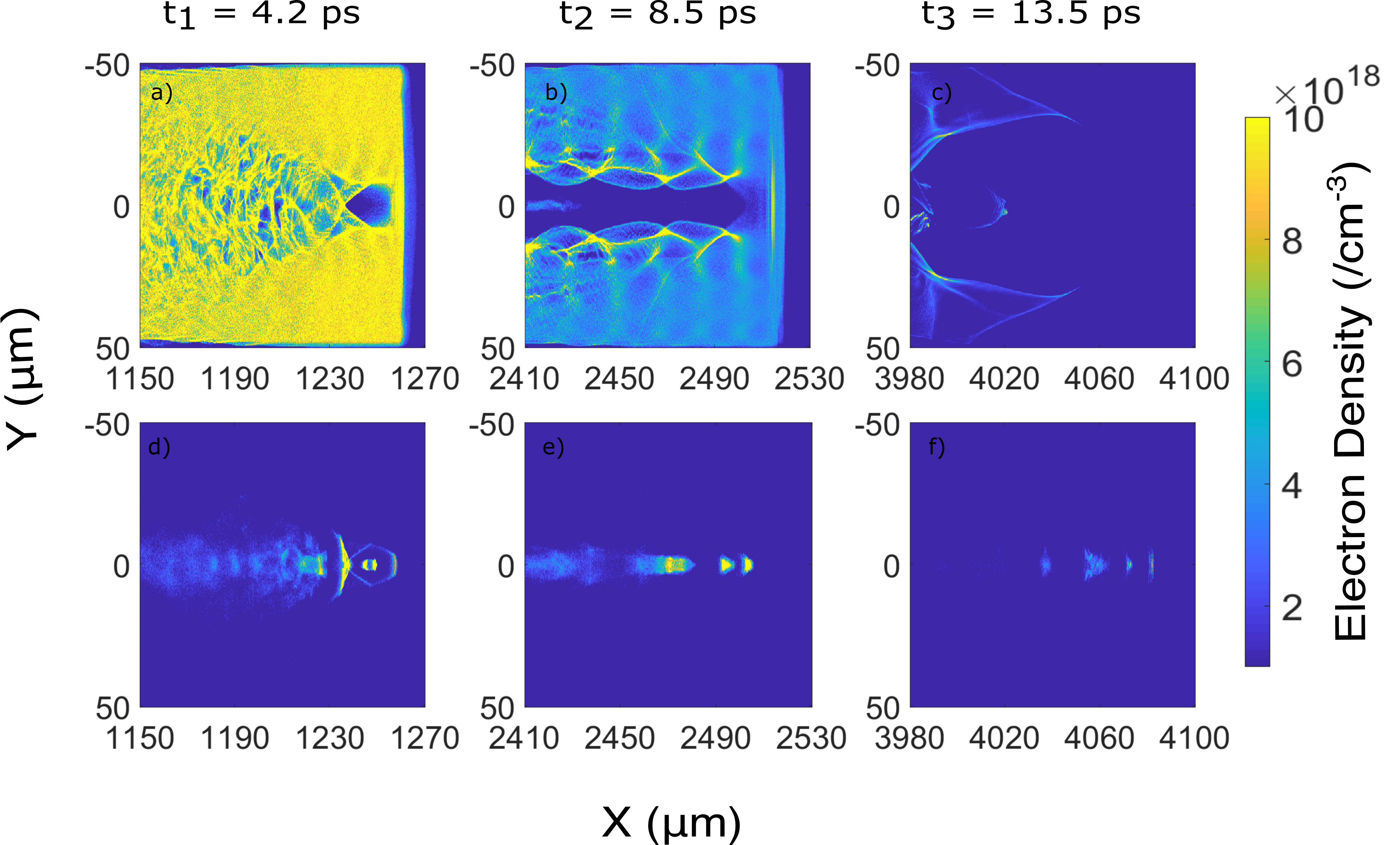}
\caption{PIC simulation results: time evolution of electron density for L-shell electrons (1-5) (up) and the 7th electron (K-shell) (down). Left frames correspond to the time that the laser crosses the peak of the plasma density, middle frames to the end of the downramp and right to the end of the simulations.}
\label{Fig.Sim}
\end{figure*}
To support our experimental results and distinguish the contribution of the injection mechanisms, we performed 2D PIC simulations. The LWFA models are simulated using EPOCH PIC code\cite{Arber_2015}. We used a moving window (moving at 0.995$\times$c), in the laboratory frame configuration. The computational domain size is of 120 $\times$ 100 \SI{}{\micro\metre}, discretized by 2048 $\times$ 384 cells. Two neutral $N$ macroparticles per cell of the 5th-order particle shape function were used \cite{Tazes_2020}. The ionization module was implemented; thus, the energy thresholds for ionization of each $N$ electron were set respectively. Given the thresholds, in the intensity regime of this work, K-shell electrons of $N$ are ionized only at the peak electric field of the laser pulse, while L-shell electrons are already ionized by the front lower-intensity region of the pulse (~\SI{e16}{W/cm^2}). To discriminate the contribution of the co-existing injection mechanisms (downramp \& ionization injection), electrons were separated into three groups, group 1 containing 5 electrons of the L-shell (already ionized before the peak electric field), and groups 2 and 3 containing 6  and 7 electrons. This approach allows for separate handling, essentially investigating the behaviour of each group. The atomic density profile has been simulated as a high-order polynomial function seen in Fig.~\ref{Fig.spec_sim}(a), which approximates well the main downramp region while taking into account the first moderate density oscillation measured in Fig.~\ref{Fig:jet}(c), avoiding using tophat and trapezoidal approximations, thus avoiding discontinuities which are far from the measured profile and enhance injection numerically. In Fig.~\ref{Fig.spec_sim}(b) the resulting spectra at the end of the simulation are presented for the sum of all K- and L-shell electrons separately. The L-shell electron population is smaller while achieving lower energy than K-shell electrons. The maximum energy approaches the maximum measured energy. Their behaviour is explained by the electron density maps seen in Fig.~\ref{Fig.Sim}. Electron density maps of L-shell electrons (top) and K-shell (bottom)  evolve at three time steps: this $t_{1}$ corresponds to the laser crossing the density peak. Fig.~\ref{Fig.Sim}(a,d), $t_{2}$ corresponds to the end of the downramp Fig.~\ref{Fig.Sim}(b,e) and $t_{3}$ at the end of the gas. Fig.~\ref{Fig.Sim}(c,f). In Fig.~\ref{Fig.Sim}(a) L-shell electrons have formed the bubble at the wake of the laser, while no electrons have yet been injected. In Fig.~\ref{Fig.Sim}(b) the bubble has evolved to a channel after continuous elongation along the downramp, and it is evident that from the rear side of the channel a bunch of electrons is injected. This population is attributed to the downramp mechanism. Finally, in Fig.~\ref{Fig.Sim}(c) accelerated electrons are released to the free space at the end of the simulation. In Fig.~\ref{Fig.Sim}(d,e) and (f), the presence of K-shell electrons is confirmed, especially close to the axis of propagation. Significantly fewer K-shell electrons are finally accelerated at the end of the simulation (than those ionised). 
Our PIC results show that the acceleration of electrons injected due to both mechanisms is confirmed.  Electrons of the K-shell  achieve energy up to \SI{285}{MeV}. The behaviour matches well with the experimental results of accelerated electrons of \SI{45}{bars}. In addition, a parametric study keeping the density profile form and varying the peak density value of the polynomial was conducted to gain more insight. By reducing density, fewer and less accelerated electrons of the L-shell are noticed in the final spectra up to the dissipation of the downramp mechanism, while in the higher density cases, downramp-injected electrons continue to accelerate efficiently. As far as K-shell electrons, their injection is confirmed for the higher- and lower-density cases. However, as the density increases, filamentation of the laser reduces the energy absorbed by the main accelerating structure and restricts their acceleration.

\section{Conclusions}

In summary, we performed an experimental study, supported by numerical simulation results, of an electron source produced by laser wakefield acceleration (LWFA), using a novel-design, 3D-printed NS nozzle. The electron source characteristics, such as the electron energy, divergence, and total beam charge, were measured at different backing pressures. The results indicate a tunable high-dose, high-energy electron source. The median total charge remained almost stable as a function of the backing pressure. The electron beam divergence remained relatively high even within the optimal backing pressure. The charge/dE/shot was enhanced by at least a factor of \SI{10}{} and the maximum energy by a factor of \SI{2}{}, compared to our previous work in which various symmetrical nozzles were implemented \cite{Grigoriadis_PPCF_2023,Grigoriadis_2022,Andrianaki}. The increase in the electron charge is attributed to the downramp mechanism, which reduces the injection threshold along with the ionization injection mechanism.
Increasing the backing pressure resulted in the acceleration of electrons with more beam-like characteristics, with the optimal operating pressure range being in the window \SI{45}{}-\SI{55}{bar}. 
Nevertheless, the low-energy electron bunches generated at low backing pressures are useful for radio-medical studies, as their energy corresponds to that of typical LINACS established in hospitals for radiotherapy (\SI{6}{}-\SI{25}{MeV}) \cite{labate_toward_2020}. Thus, comparative experiments can be performed to explore the potential advantages of pulsed electron sources in radiotherapy \cite{fitilis22}. In addition, the dose profile shows a central iso-surface with significant dose accumulation. Finally, the resulting electron bunches are considered suitable for Betatron radiation emission (i.e., pump-probe experiments) \cite{APL_Betatron_nitrogen,Grigoriadis_2022}, as this radiation is enhanced as a function of electron charge; thus, high X-ray flux is expected \cite{Ferri_betatron, Emma_Van_2021,Kamperidis_Papp_2021}. 

\begin{acknowledgments}
The author D.M. acknowledges funding by the Hellenic Mediterranean University within the project "Proposal for post-doctoral research at the Institute of Plasma Physics and Lasers (IPPL) of Hellenic Mediterranean University (HMU)" on the context of the 2607/$\Phi$.120/04-05-2022 call of HMU for post-doctoral research. 
This work was carried out within the project “Use of secondary sources of plasma radiation produced by the interaction of powerful lasers with material for biomedical applications” funded by national resources of Greece under the Regional Development Program of the Region of Crete (2024N$\Pi$10200000/5220198). 

The EPOCH code used in this work was in part funded by the UK EPSRC grants EP/G054950/1, EP/G056803/1, EP/G055165/1, EP/ M022463/1 and EP/P02212X/1. 
We acknowledge the support with computational time granted by the Greek Research \& Technology Network (GRNET) in the National HPC facility ARIS – under project ID pr016025-LaMPIOS III.

This work has been carried out within the framework of the EUROfusion Consortium, funded by the European Union via the Euratom Research and Training Programme (Grant Agreement No 101052200 — EUROfusion) and the Hellenic National Program of Controlled Thermonuclear Fusion. Views and opinions expressed are however those of the authors alone and do not necessarily reflect those of the European Union or the European Commission. Neither the European Union nor the European Commission can be held responsible for them.

\end{acknowledgments}

\section*{Data availability}

The data that support the findings of this study are available from the corresponding author upon reasonable request.

\section*{References}
\bibliography{aipsamp}

\end{document}